\documentclass[12pt]{article}
\usepackage{graphicx}
 \begin{document}

\begin{figure}[ht]
\begin{center}
\[
\mbox{\begin{picture}(200,350)(60,40)
\includegraphics[scale=.25]{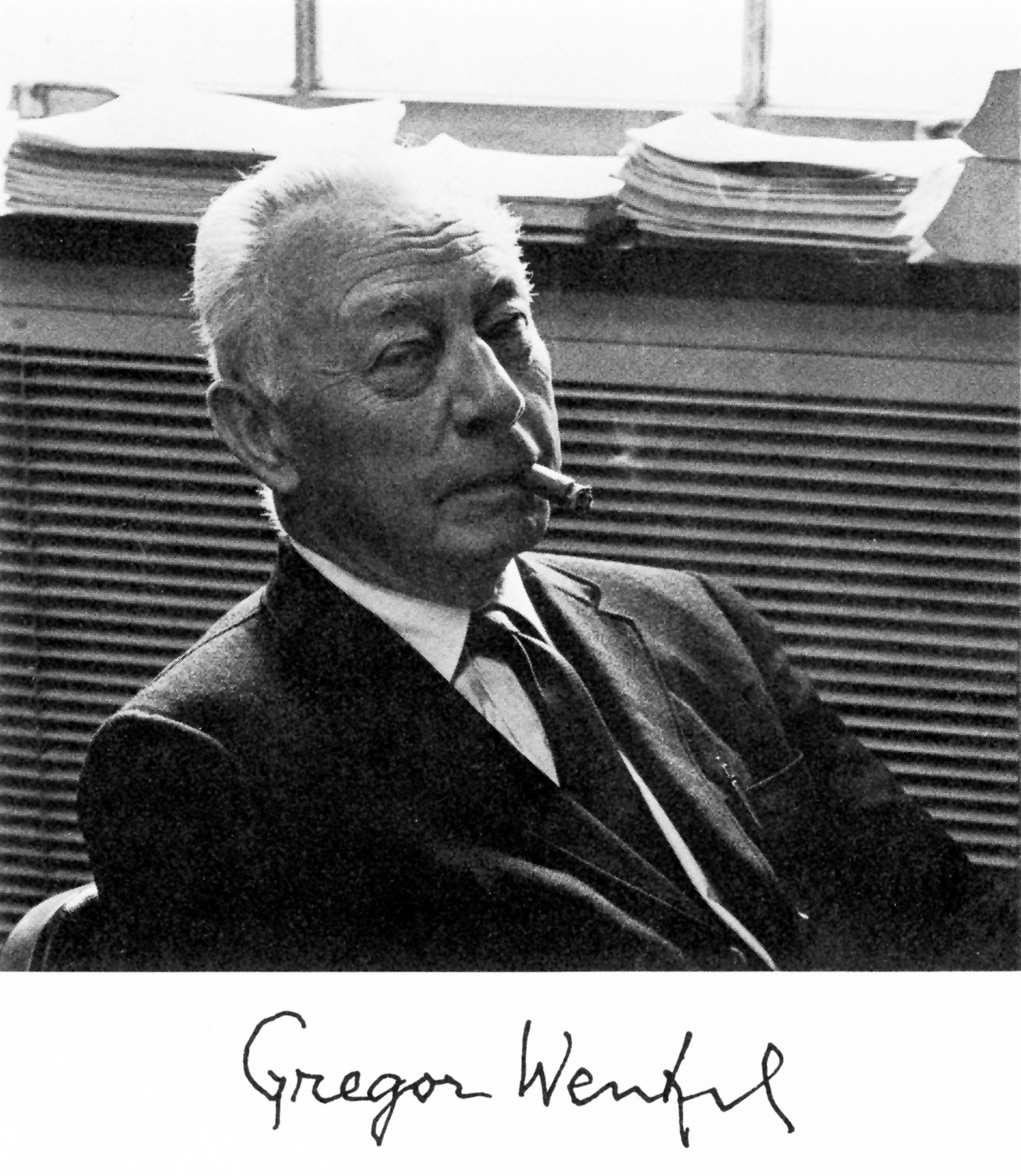}
\end{picture}}
\]
%
\end{center}
\end{figure}

\medskip
\begin{center} \large
               {\bf GREGOR WENTZEL} \\
	\normalsize \medskip
	{\em February 17, 1898 - August 12, 1978} \\
	\bigskip\bigskip
	BY PETER G.O. FREUND, CHARLES J. GOEBEL, \\ YOICHIRO NAMBU AND REINHARD OEHME  	   

 \end{center}

\bigskip\bigskip \bigskip
\vfill\eject

{\bf  Biographical sketch}
\medskip

The initial of Gregor Wentzel's last name has found a solid place in the language of theoretical physics as the W 
of the fundamental WKB approximation, which describes the semi-classical limit of any quantum system. Beyond this 
fundamental contribution to quantum theory \cite{W1}, Gregor Wentzel has played an important role in the theoretical 
physics of the first half of the twentieth century.

Born in D\"{u}sseldorf on 17 February, 1898, Gregor benefited from a rich and multi-faceted education. As the greatest 
events of his youth, Wentzel used to recall the local premi\`{e}res of the symphonies of Gustav Mahler. A lifelong love 
for music was instilled in the young Gregor. During World War I he served in the army from 1917 to 1918. At the 
conclusion of that cataclysmic event, he continued his studies, migrating from university to university, as was 
customary in those days. First, until 1919 we find him at the University of Freiburg, then at the University of 
Greifswald, and as of 1920, just like Wolfgang Pauli and Werner Heisenberg and then later Hans Bethe among others, 
studying with 
the legendary teacher Arnold Sommerfeld at the Ludwig Maximilians University in Munich, where he obtained his Ph.D. 
with a thesis on Roentgen spectra \cite{W2}. Still in Munich he completed his {\em Habilitation} in 1922 and became 
a {\em Privatdozent} (roughly the equivalent of what today would be an assistant professor). In 1926 Wentzel moves to 
the University of Leipzig as an {\em a.o. Professor} (roughly an associate professor). Then, in 1928, he is 
appointed Erwin Schr\"{o}dinger's successor as professor of physics at the University of Zurich. That same year Wolfgang 
Pauli takes over the chair of theoretical physics at the Federal Institute of Technology (ETH) in Zurich. These two 
former Sommerfeld students become the joint leaders of physics in Zurich, one of the German-speaking world's most 
vibrant scientific communities.

During World War II, Pauli --- a three-quarters Jewish citizen of the Third Reich after the annexation of his 
native Austria --- unable to obtain Swiss citizenship, flees to the United States and Wentzel remains in Zurich to 
single-handedly see to the maintenance of the high standards of theoretical physics research and teaching there. 
After the war, Pauli returns to Zurich and Wentzel is offered a professorship at the University of Chicago. He moves 
to Chicago in 1948 and remains there until his retirement in 1970. The Wentzels then move to Ascona, 
Switzerland. For his contributions to theoretical Physics the German Physical Society awards Gregor Wentzel the 1975 
Max Planck medal, its highest honor. The octogenarian Gregor Wentzel dies in Ascona on 12 August 1978, survived 
by his wife Annie 
and his son Donat G. Wentzel, currently a professor emeritus of astronomy at the University of Maryland.

\bigskip \bigskip

{\bf  Scientific Biography}
\medskip

Gregor Wentzel's scientific life neatly divides into three well-defined periods:

1) His work on erecting the glorious edifice of Quantum Mechanics, 

2) His work on meson theory, and

3) His work on condensed matter Physics.

\bigskip

We start with his work on Quantum Mechanics. 
The old quantum "theory" taught by Sommerfeld was too successful to be abandoned altogether and too riddled with 
contradictions to be able to survive for long. 1925 was the decisive year during which Heisenberg discovered matrix 
mechanics \cite{H}, and soon thereafter Schr\"{o}dinger \cite{S} produced his famous equation. From two decades of 
work with the old theory, many of the crucial questions were already clearly in place. 

One of these questions asked how 
the transition from the quantum regime to the classical regime is achieved, in other words, how does one 
calculate the lowest-order 
quantum corrections. This question was answered in three now classical papers by Gregor Wentzel \cite{W1}, Hendrik 
Anthony Kramers \cite{K} and L\'{e}on Brillouin \cite{BR}. 

In a brilliant piece of mathematical physics, 
Wentzel recasts the simplest 1-dimensional Schr\"{o}dinger equation, 
as a Riccati equation. He then expands the function which appears in this Riccati equation in a power series in Planck's 
constant, and is led to a set 
of recursion relations for the coefficient functions in this series expansion. These coefficient functions describe 
the quantum corrections. Wentzel then successfully 
applies this approximation method to the Hydrogen atom and to the Stark effect.
 
Brillouin's independent paper \cite{BR}, presented on July 29, 1926 to the Paris Academy is along the same lines as 
Wentzel's work, submitted one month earlier (June 18, 1926) to the journal {\em Zeitschrift f\"{u}r Physik}. Kramers' 
later work \cite{K} is based on saddle point and steepest descent techniques, and provides the famous discussion 
of turning points. This approximation method was known in various mathematical contexts,, see \cite{J,L,C} among 
others. The great importance of its independent discovery by WKB consists in the fact that, when 
applied to the then new 
Schr\"{o}dinger equation, this approximation scheme solves 
the major physical problem of systematically calculating quantum corrections.

The next problem tackled by Wentzel was that of the photoeffect, the phenomenon, which, when first studied by Albert 
Einstein \cite{E} led to the introduction of the light quantum concept, and thus played a pivotal role in the 
construction of quantum theory. Wentzel \cite{W3}, and independently P.A.M. Dirac \cite{D}, were the first to 
give a full fledged quantum-theoretic treatment of the photoeffect. In \cite{W3} Wentzel finds the angular 
distribution of the photoelectrons. Wentzel then derives the intensity of the photoemission by writing 
down the formula  known nowadays as Fermi's golden rule, given the subsequent extensive use and emphasis placed on it 
by Enrico Fermi.

In 1926 Max Born published his celebrated paper \cite{B1} about the application of quantum-mechanical perturbation 
theory to scattering processes, the so-called Born approximation. In this paper Born introduced the first explicit 
statement of the probabilistic interpretation of quantum mechanics. Wentzel often used to remark that "Born did not 
choose to write a separate paper on the probabilistic interpretation, because at that point this interpretation, 
though never spelled out in print, was known to everyone working in the field." The obvious application of Born's 
method would have been to the Coulomb problem. There was the difficulty that a direct application did not yield a finite 
result, on account of the divergence of Born's integral expression in this case. 
What Wentzel did \cite{W4}, was simply to
provide the Coulomb potential with an exponential factor (thus turning it into what later would become known as a 
Yukawa potential), which renders Born's integral convergent. He then carried out the Born integral and {\em in the 
end result} got rid of the exponential by letting the coefficient in its exponent vanish. This famously leads 
to the well-known Rutherford formula.

It is rarely mentioned in the literature, that we owe this result to Wentzel. Born himself was aware of Gregor's 
paper and in his Nobel lecture \cite{B2} he quotes it as, "Soon Wentzel succeeded in deriving Rutherford's
famous formula for the scattering of $\alpha$-particles from my theory." Wentzel's take: "Born was too mathematical 
to dare alter the Coulomb potential. I had no such compunctions and Born never forgave me for that." It is amusing 
that the paper \cite{W4} on the Rutherford formula and the paper on the photoeffect, \cite{W3}, two of Wentzel's 
three best known papers from this period, were submitted for publication on the same day, and appear next to each 
other in {\em Zeitschrift f\"{u}r Physik}.

In fact, Wentzel's interest in scattering processes predates Quantum Mechanics. Already in the old quantum theory he  
did the most advanced analysis of scattering processes \cite{W5}. Some of the results derived by him 
have remained in use even 
after the discovery of quantum theory. Not surprisingly, the first comprehensive review of 
scattering and radiation processes is due to Wentzel \cite{W6}

We must mention here that in a prophetic 1924 paper \cite{W7} Wentzel attempted a "sum over paths" approach 
to the construction 
of quantum amplitudes. 
He weighted each term in the sum by the nowadays well-known quantum phase factor, which accounts for quantum 
interference phenomena. From the so obtained quantum amplitudes he then obtained probabilities by taking the 
squares of their absolute value.

\bigskip 
The second phase of Wentzel's work deals with Yukawa's then new meson theory, its highlight being Wentzel's 
famous strong coupling approximation.

\bigskip

Wentzel's first publication on meson theory was a review \cite{W8} describing Yukawa's suggestion \cite{Y} that 
the nuclear force could be due to a field whose quanta would be bosons a few hundred times more massive than 
the electron, and the demonstration by Neddermeyer and Anderson \cite{NA} that particles of such mass constituted 
the so-called penetrating part of cosmic rays.  Soon after this article it became clear that the observed particles 
did not have the strong nuclear interactions expected of Yukawa's particle. This serious difficulty for meson 
theory was not overcome until 1947, when experiments showed that the bulk of the penetrating cosmic rays 
came in the form of a totally unexpected particle, the mu-lepton, itself a decay product of the charged Yukawa mesons.

Wentzel initiated the strong coupling approximation to the static meson model (`static' here means 
neglect of nucleon motion) in his 1940 article \cite{W9} in which, as suggested by Yukawa, he treated the 
simplest case of a spinless charged meson field, with an s-wave coupling to nucleons.  He begins by saying that the 
usual perturbation theory 
which expands in powers of the meson-nucleon coupling constant, $g$, cannot be used, since the observed strength of 
he nuclear force implies that $g$ is large.  Wentzel therefore turns to the alternative of an expansion in the 
reciprocal of $g$ and finds that

a) for large $g$, the nucleon has low-lying multiply charged excitations, 
{\em isobars}, interpretable either as bound states of a nucleon and mesons or as rotational states (in an internal 
`charge space') of a rigid rotator, and 

b) the meson-nucleon scattering cross section does not increase without 
limit for increasing $g$.  Of course, unitarity (probability conservation) enforces an upper bound on the 
cross-section
in any scattering calculation, if done 
sufficiently correctly; Wentzel's seems to be the first which was. 

Wentzel says that unfortunately, this limiting cross section is at least a hundred times what had been observed.  
For only moderately large $g$ the cross section will of course be less, but he suspects that the original 
Yukawa theory is in trouble, and perhaps one should try a spin 1 meson

It soon became clear that the properties of the nuclear force required the Yukawa meson to have i-spin 1 
(so-called charge-symmetric theory) and to be p-wave coupled to the nucleon spin, so mesons had to be pseudoscalar 
or vector (spin$^{parity}$ $0^-$ or $1^-$).  The strong coupling calculation for this kind of meson field 
was first published by Pauli and Dancoff \cite{PD}.
It was followed by a 1943 elaboration by Wentzel in Zurich \cite{W10}.
  The isobars were found to have all possible half-odd-integer values  of both spin $j$ and i-spin $i$, 
  with the remarkable restriction $j=i$ (this same feature occurs in the Skyrme 
  model \cite{SK} of the nucleon as a topological soliton of a simple non-linear equation for the meson field).  
  Thus the ground state has $j\!=\!i\!=\!\frac{1}{2}$ and represents the proton and neutron. The first 
  excited state is predicted to have $j\!=\!i\!=\!\frac{3}{2}$.  
  
  A decade later Brueckner (1952) pointed out that the 
  ongoing Chicago synchrocyclotron $\pi$-proton scattering results were well fitted by a $j\!=\!i\!=\!\frac{3}{2}$ 
  resonance peak at $\approx \! 200$ Mev, now known as the $\Delta(1238)$ baryon. 
  It had always been assumed in 
  strong coupling theory that isobars had to be bound states, i.e. unable to decay with emission of a meson. Yet 
  for p-wave 
  isobars the bound/unbound distinction is blurred by centrifugal barrier decay suppression. This may explain why 
  no connection was even suggested, whether by Wentzel himself or by anyone else, between 
  the experimentally diecovered $\Delta(1238)$ baryon and the isobar predicted by the strong coupling calculation. 
  It should be 
  added, however, that the next isobar, $j\!=\!i\!=\!\frac{5}{2}$, has never been found. In the quark model its absence 
 from the spectrum is accounted for, since such a baryon  
  cannot be a three-quark state.

In 1947 Wentzel wrote another review \cite{W11} on meson theory, discussing strong coupling and other
attempts to get a sufficiently small meson-nucleon cross section, just before this problem evaporated 
with the discovery by Lattes, Muirhead, Occhialini and Powell \cite{LMOP} that the penetrating cosmic ray 
particles were only a decay 
product of what they named the pi meson (`p' for `primary'), the actual Yukawa meson.  Shortly thereafter 
pi-mesons were produced at the Berkeley cyclotron \cite{GL}, which allowed rapid determination of their 
basic properties.

In his later years, Wentzel made strong coupling calculations \cite{W12, W13, W14} for K mesons scattering 
on $\Lambda$- and $\Sigma$-hyperons.  Some of these exhibited interesting features such as a rapid switch 
of the nature of the clothed hyperon (the rigid rotator) at a critical value of the ratio of $\Lambda$-coupling 
to $\Sigma$-coupling, or a higher symmetry than the Hamiltonian had. However, the results did not have much resemblance 
to observation.

\bigskip

The third phase of Wentzel's work is devoted to condensed matter physics and many-body problems. Wentzel's 
activity in these fields  started in Chicago, in his later years after the war. Considerable 
progress in the area took place in the 1950s, especially in understanding superfluidity 
and superconductivity from first principles. So it was natural for Wentzel to get involved. He published several 
papers \cite{W15, W16, W17} during this decade, although they are not among his main contributions to physics. 
Basically his role 
was that of promoter and critic.
Wentzel appreciated exact results obtained by others, for example, in the properties of an electron gas with 
interaction, and used his familiarity with field theoretical techniques to generalize or simplify them, at the 
same time criticizing deficiencies in some papers. In particular, he took great interest in the BCS theory \cite{BCS} 
of superconductivity and the associated Bogoliubov-Valatin \cite{B,V} description 
of electrons (quasiparticles). Like many physicists, 
however, Wentzel was not satisfied with the lack of gauge invariance in the theory. In a paper on the Meissner 
effect \cite{W18} he proposed a modification of the BCS procedure, but a better understanding of this issue was 
left to the developments concerning spontaneous symmetry breaking and mass generation for gauge fields.

\bigskip \bigskip

{\bf Gregor Wentzel, teacher and colleague}
\medskip

Wentzel's lectures starting in his early Leipzig days, all the way to the memorable courses he taught 
at the University of Chicago, have always awed listeners by their exquisite elegance. This quality is borne out
in his textbook {\em Einf\"{u}hrung in die Quantentheorie der Wellenfelder} \cite{W19}, written 
during the war. This first book ever on quantum field theory was translated into English at war's end \cite{W20}, and 
has been the formative textbook of the postwar generation of theoretical physicists.

Wentzel's list of doctoral students is truly remarkable. On it we find
Valentin Bargmann, Markus Fierz, Res Jost, Nicholas Kemmer, 
A. Houriet, Felix Villars, Fritz Coester, Josef Maria Jauch, Burton Fried, Allan Kaufman, 
Charles Goebel, Nina Byers and R. Ramachandran. 

In a sense, to this list of former Wentzel students, one could add the name of Homi J. Bhabha. 
A student at Cambridge, Bhabha, who later became one of the first internationally recognized Indian 
theoretical physicists, had been dispatched by his adviser to work with Pauli in Zurich. After reading the adviser's
not all that good 
letter of introduction, Pauli refused to have anything to do with Bhabha. It is though in Zurich, 
in this hostile atmosphere, that Bhabha wrote his most 
famous paper, the paper on electron-positron scattering, or "Bhabha scattering" as it is now called. 
When he tried to show it 
to the great man, Pauli responded "If {\em you} did this, I am not interested." At wit's end, 
Bhabha went to Gregor, the other senior theoretical physicist in town.  Familiar with his good friend 
Pauli's quirks, Gregor took the paper and read it. He immediately realized its importance and assured its young author 
that he would convince Pauli of its merit. At first Wentzel's attempt to explain Bhabha's work to Pauli 
was met with the expected "If Mr. Bhabha did it, I am not interested," but Gregor was prepared for this. "For just 
a moment" he suggested, "imagine that I had done it." Pauli was willing to listen. Bhabha remained forever in  
Wentzel's debt.

Beyond teaching, a professor also has to conduct exams. Wentzel had a remarkable technique for doing this. 
He would start with a question about a simple system, say the classical non-relativistic rotator. Then he would 
build up from this to a slightly more complex setting by asking for a quantum mechanical description, then
moving up to the relativistic case. Sooner or later, the student couldn't answer his question, and where this
breakpoint was reached determined the student's grade.

At the University of Chicago, just like earlier in Zurich, Wentzel was a
central figure among theoretical physicists, able to create a marvelous scientific atmosphere.
First of all there were the frequent and unscheduled private meetings 
in Gregor's office. Two of us (YN and RO) came to Chicago in time to still catch some of these meetings.
Surrounded by the aromatic smoke of Wentzel's ever-present
Cuban cigars, stimulating discussions on the latest problems were held in a
completely informal setting. Maria Goeppert-Mayer would report
on her shell-model calculations, Gregor himself on strong-coupling meson theory and
on hyperon decays and Enrico Fermi 
on the results obtained in his pion-nucleon scattering experiments at
the in-house synchro-cyclotron. Reinhard Oehme presented his results on the connection by analytic
continuation of particle-particle and antiparticle-particle scattering
amplitudes, and on the corresponding dispersion relations for
pion-nucleon scattering then freshly obtained by
him, Murph Goldberger and Hironari Miyazawa.
These dispersion relations agreed nicely with the experiments carried out by Fermi's group. 
In his own relaxed, spontaneous and inimitable way Gregor moderated these discussions, and kept asking 
pertinent questions. Sometimes
he would invite outsiders --- a visit from Hans Bethe comes to mind --- in order to explain
points in their work.

At the Fermi Institute, or Institute for Nuclear Studies, as it was then called, there was a weekly seminar, where
the Institute's members presented their results. In the beginning Enrico Fermi, Willard Libby, Harold Urey and Gregor 
Wentzel jointly led this seminar. Gell-Mann's first talk ever about strangeness was presented there. 
With Fermi's death in 1954 and the subsequent departure
from Chicago of Libby and Urey, Wentzel became the seminar's single leader. 

It was like no other seminar or colloquium.
There was a fixed time for it, Thursdays an hour and a quarter 
before the Department of Physics colloquium, but beyond that
absolutely nothing was planned. You went to this seminar prepared to talk for anywhere between ten and thirty minutes
on what you were working, if and only if called upon.
As the seminar started, Wentzel would turn around in his front-row seat, look at one of the attendants and invite him
(at the risk of admitting to political incorrectness, after Maria Goeppert-Mayer's departure from Chicago, 
there were no women in attendance in those ancient times) to speak. 
This was before the age of the laptop and even of the transparency, and people had to make do with chalk and
blackboard, and being spontaneous, they had no notes along either. It was the most exciting seminar
many of us ever attended. Chandrasekhar often reported the sensational news about quasars. As the 
Astrophysical Journal's editor-in-chief, he knew 
about these new spectacular findings long before they were made public. 
Very excited, he prefaced 
each report with a demand of confidentiality; he swore the audience to silence, as it were. It was 
high drama at the very frontier of science.

\medskip

Gregor was a modest man, and at all times very much the gentleman. He made his mark on Physics both through 
his own important work and through his legacy as a teacher. This legacy, not unlike that
of his own teacher, Arnold Sommerfeld, is remarkable for the brilliant physicists it helped 
shape. It is also remarkable for the great role Wentzel's book has had in setting the direction of postwar research in
theoretical Physics.

\end{document}